\documentclass[aps,prd,twocolumn,superscriptaddress,showpacs,showkeys]{revtex4-1}  
\usepackage{graphicx,color} 
\usepackage{xcolor}
\usepackage[hidelinks]{hyperref}
\usepackage{rotating}
\usepackage{amssymb}
\usepackage{amsfonts}
\usepackage{amsmath}
\usepackage{xspace}
\usepackage{mathtools} 
\usepackage{verbatim}

\DeclarePairedDelimiterX\Braket[2]{\langle}{\rangle}{#1 \delimsize\vert #2}

\usepackage{listings}
\makeindex

\usepackage[thinlines]{easytable}

\usepackage[super]{nth}

\newcommand{\be}{\begin{equation}}
\newcommand{\ee}{\end{equation}}
\newcommand{\bea}{\begin{eqnarray}}
\newcommand{\eea}{\end{eqnarray}}

\newcommand\ie{\mbox{\textit{i.\,e.}}\xspace}

\newcommand\eg{\mbox{e.\,g.}\xspace}

\newcommand\D{\mathrm{d}}

\newcommand{\hx}{\hat{x}}
\newcommand{\hp}{\hat{p}}
\newcommand{\hX}{\hat{X}}
\newcommand{\hP}{\hat{P}}
\newcommand{\mpar}{\dot{\partial}}

\begin{document}

\title{Generalized uncertainty principle or curved momentum space?}




\author{Fabian Wagner}
\email[]{fabian.wagner@usz.edu.pl}
\affiliation{Institute of Physics, University of Szczecin, Wielkopolska 15, 70-451 Szczecin, Poland}
\date{\today}
\begin{abstract}
The concept of minimum length, widely accepted as a low-energy effect of quantum gravity, manifests itself in quantum mechanics through generalized uncertainty principles. Curved momentum space, on the other hand, is at the heart of similar applications such as doubly special relativity. We introduce a duality between theories yielding generalized uncertainty principles and quantum mechanics on nontrivial momentum space. In particular, we find canonically conjugate variables which map the former into the latter. In that vein, we explicitly derive the \emph{vielbein} corresponding to a generic generalized uncertainty principle in $d$ dimensions. Assuming the predominantly used quadratic form of the modification, the curvature tensor in momentum space is proportional to the noncommutativity of the coordinates in the modified Heisenberg algebra. Yet, the metric is non-Euclidean even in the flat case corresponding to commutative space, because the resulting momentum basis is noncanonical. These insights are used to constrain the curvature and the deviation from the canonical basis. 
\end{abstract}

\pacs{}
\keywords{}

\maketitle

\section{Introduction}
\label{sec:intro}

The idea of a fundamental limitation to length measurements, originally going back to work of Werner Heisenberg \cite{Heisenberg30} and Hartland Snyder \cite{Snyder46} and encountered in string theory \cite{Amati87,Gross87a,Gross87b,Amati88,Konishi89}, loop quantum gravity \cite{Maggiore93b,Hossain10,Majumder12a,Girelli12,Gorji15}, noncommutative geometry \cite{Battisti08c,Pramanik13} as well as Ho\v{r}ava-Lifshitz gravity \cite{Myung09b,Myung09c,Eune10} but also derived from general arguments combining gravity and quantum theory \cite{Mead64,Mead66,Padmanabhan87,Ng93,Maggiore93a,Amelino-Camelia94,Garay94,Adler99a,Scardigli99,Capozziello99,Camacho02,Calmet04,Ghosh10,Casadio13,Casadio15,El-Nabulsi20a} since, has played a prominent r\^ ole in the literature on the phenomenology of quantum gravity. In quantum mechanics, such a minimum length may be implemented by invoking a generalized uncertainty principle (GUP) which, in turn, may be derived from a momentum-dependent deformation of the Heisenberg algebra \cite{Maggiore93c,Kempf94,Kempf96a,Benczik02,Das12}.

Over the last 30 years this approach has continuously gained momentum in the community leading to manifold applications \cite{Bawaj14,Bushev19,Das11,Ghosh13,Girdhar20,Das08,Bouaziz10,AntonacciOakes13,Marin13,Marin14,Das08,Ali11b,Gao16,Khodadi18a,Petruzziello20,Buoninfante20}. Note, however, that it harbours a number of subtleties, in spite of the success, many of which are carefully reviewed in Ref. \cite{Hossenfelder12}. For example, it suffers an inverse soccer problem rooted in the fact that the corrections to the dynamical variables of the center of mass in multiparticle states are inversely proportional to the number of constituents \cite{Amelino-Camelia13}. This begs the question what a fundamental constituent is supposed to be. Furthermore, the deformed commutator can only yield either a trivial or a divergent classical limit \cite{Casadio20} implying that it is a purely quantum mechanical effect \cite{Chashchina19}. This just closely saves it from violating Gromov's non-squeezing theorem \cite{Gromov85}, a hallmark of symplectic geometry which may be understood as classical analog of Heisenberg's uncertainty principle \cite{deGosson09}. In that vein, the GUP may also challenge the second law of thermodynamics \cite{Hanggi13}. Moreover, its synthesis with the principle of gauge invariance is not thoroughly understood \cite{Chang16} and its relativistic extensions lead to deformations \cite{Ali09} or straight violations \cite{Lambiase17} of Lorentz invariance. Last but not least, as was alluded to above, the minimum length may be derived from high-energy string scattering amplitudes \cite{Amati87,Amati88}. However, its value differs from the one inferred from D-branes \cite{Shenker95,Douglas96} making the GUP probe-dependent in string theory.

Curved momentum space, in contrast, is a very old idea which is only gradually attracting attention within physics. The first record of it in mathematics dates back to Bernhard Riemann's habilitation dissertation \cite{Riemann54}. Later, it was mainly developed by Paul Finsler \cite{Finsler18} and \'Ellie Cartan \cite{Cartan34}. An overview of this topic, these days subsumed under the terms Lagrangian and Hamiltonian geometry, can be found in Refs. \cite{Miron01,Miron12}.

Conceived by Max Born \cite{Born38} as a necessary condition for the generalization of the symmetry of flat-space mechanics under the exchange of phase space variables $\hat{x}\rightarrow\hat{p},$ $\hat{p}\rightarrow -\hat{x},$ nowadays called Born reciprocity, to curved spacetime, nontrivial momentum space was intended to pave the way towards a unification of quantum theory and general relativity. This approach was further developed mainly by Yuri Gol'fand \cite{Golfand59,Golfand62,Golfand63} and Igor Tamm \cite{Tamm65,Tamm72}. From the mathematical side said endeavour lead to the theory of quantum groups \cite{Drinfeld88,Majid88,Majid90,Majid91,Freidel05}. Furthermore, the canonical quantization of theories on curved momentum space was treated in Refs. \cite{Batalin89a,Batalin89b,Bars10}. These efforts culminated in their recent application to quantum gravity phenomenology \cite{Amelino-Camelia11,Carmona19,Relancio20a} on the one hand. On the other hand, they paved the way for the construction of Born geometry \cite{Freidel13,Freidel14,Freidel15,Freidel17,Freidel18}, which captures all mathematical structures behind Hamiltonian mechanics (symplectic), quantum theory (complex) and general relativity (metric) at once.

A deviation from ordinary quantum mechanics, analogous to the GUP but deduced from position-dependent corrections, goes under the name extended uncertainty principle \cite{Bambi08a,Mignemi09,Ghosh09,CostaFilho16}. Recently, such a relation was derived from curved position space alone \cite{Schuermann18,Dabrowski19,Dabrowski20,Petruzziello21}. However, the generality of the arguments provided there allows for a reinterpretation: As argued in Ref. \cite{Petruzziello21}, following the same reasoning while taking curved momentum space as the starting point, it would be possible to obtain a GUP without assuming a deformation of the algebra of observables. Interestingly, this kind of link had already been studied in Ref. \cite{Chang10}. The connection between curved spaces and modified algebras is on display in the context of doubly special relativity \cite{Amelino-Camelia02b,Kowalski-Glikman02b,Amelino-Camelia02c,Amelino-Camelia02c,Amelino-Camelia10} as well, which may be interpreted as a theory defined on de Sitter-momentum space \cite{Kowalski-Glikman02c,Kowalski-Glikman03}. Moreover it has been corroborated further from the geometric point of view \cite{Carmona21,Relancio21}. Those results provide a strong motivation to search for an equivalence between GUP-deformed quantum mechanics and quantum mechanics on curved momentum space. 

The aim of the present paper lies in establishing such an equivalence by introducing a novel set of conjugate variables $\hat{X}^i$ and $\hat{P}_i$ satisfying the $d$-dimensional Heisenberg algebra. Those can be used to describe these kinds of modifications in $d$ dimensions canonically. As for the transformation often applied in case of the GUP on commutative space \cite{Brau06,Bosso18b}, this naturally leads to a modification of the single-particle Hamiltonian. The thus arising dynamics constitute motion on a nontrivial momentum space. For the quadratic GUP the curvature tensor is proportional to the noncommutativity of space. However, a commutative space does not imply that the corresponding background is trivial. To the contrary, the resulting basis in momentum space is nonlinearly related to the one underlying the Euclidean metric.

Therefore, it is possible to import bounds on the curvature of momentum space and the deviation from the canonical basis from the literature on noncommutative geometry and GUPs on commutative space, respectively. We thus obtain a distinct interpretation for the already existing phenomenology. Furthermore, the new set of phase space variables allows for a rather simple treatment of noncommutative space in quantum mechanics mapping it onto a theory which is analogous to quantum mechanics on curved manifolds as described in Ref. \cite{DeWitt52}. Note that an instance of this duality was obtained along a complementary road \cite{Singh21} during the review process of the present work.

This paper is organized as follows. Sections \ref{sec:curvedmom} and \ref{sec:GUPQM} serve as brief introductions to the influence of curved momentum space and GUP-like deformations on quantum mechanics, respectively. The equivalence of those two theories is established in section \ref{sec:equivalence} providing the map connecting them. Subsequently, the newly appearing geometrical observables are constrained in section \ref{sec:constraints}. Finally, section \ref{sec:Conclusion} is intended as summary and conclusion of the results.

\section{Curved momentum space}\label{sec:curvedmom}

In order to understand curved momentum space, a short introduction to the geometry of generalized Hamilton spaces is indispensable. On the base of this reasoning and under the assumption that the metric bear no position-dependence, it is straight-forward to construct the corresponding quantum theory.

\subsection{Geometry}\label{subsec:geo}

The theory of curved momentum spaces derives from the geometry of generalized Hamilton spaces \cite{Miron01,Miron12} which is gradually seeing more application to physics, in particular in the context of the phenomenology of quantum gravity \cite{Barcaroli15,Carmona19,Relancio20a}. The starting point for this investigation is a metric which not only depends on the position but also the momentum of the investigated object
\begin{equation}
    g^{ij}=g^{ij}(x,p).
\end{equation}
To investigate the corresponding geometry, it is necessary to find a nonlinear connection $N_{ij}$ which governs the division of the cotangent bundle into horizontal ("position") and vertical ("momentum") space. This choice is highly nontrivial, though it can be simplified in a special case. Define the Cartan tensor of the background space as
\begin{equation}
    C^{kij}=\frac{1}{2}\mpar^kg^{ij},\label{cartan}
\end{equation}
where the partial derivative with respect to momenta is denoted as $\mpar^i=\partial/\partial p_i.$ If this tensor turns out to be totally symmetric, the metric can be derived from the Hamiltonian of a free particle of mass $m$
\begin{equation}
    H=\frac{1}{2m}p_ip_jg^{ij}
\end{equation}
according to the relation
\begin{equation}
    g^{ij}=m\mpar^i\mpar^jH.
\end{equation}
Furthermore, a canonical nonlinear connection can be found as
\begin{equation}
    N_{ij}=\frac{1}{4}\left(\left\{g_{ij},H\right\}-g_{ik}\mpar^k\partial_jH-g_{jk}\mpar^k\partial_iH\right),
\end{equation}
where the symbols $\{,\}$ denote the Poisson bracket. Once the nonlinear connection is known, it is possible to derive the covariant derivatives in position and momentum space and the curvature tensors. 

Assuming that the metric be solely a function of the momenta
\begin{equation}
    g^{ij}=g^{ij}(p),\label{mommet}
\end{equation}
the nonlinear connection immediately vanishes making the problem particularly simple. Correspondingly, the covariant derivative in position space is just the partial derivative. Motion in momentum space, on the other hand, is described by a Levi-Civita-like connection related to the Cartan tensor
\begin{equation}
    C^{ij}_k=-\frac{1}{2}g_{kl}\left(\mpar^{i}g^{jl}+\mpar^{j}g^{il}-\mpar^{l}g^{ij}\right).
\end{equation}
Defining covariant differentiation in momentum space denoted by the symbol $\dot{\nabla}$ in the usual way, this makes it possible to construct a scalar from the Cartan tensor 
\begin{equation}
    C\equiv g_{(jk}g_{il)}\dot{\nabla}^lC^{ijk},\label{cartsca}
\end{equation}
where the parenthesis in the indices implies total symmetrization. If the Cartan tensor is totally symmetric, this quantity is uniquely defined and measures the departure from Riemannian geometry. Moreover, the curvature tensor in position space vanishes while its counterpart in momentum space $S_k^{~ilj}$ takes the familiar form
\begin{equation}
    S_k^{~ilj}=\mpar^jC_k^{il}-\mpar^lC_k^{ij}+C_k^{ml}C_m^{ij}-C_k^{mj}C_m^{il},\label{MomRiem}
\end{equation}
which is clearly reminiscent of the Riemann tensor. Therefore, the Hamilton geometry derived from a purely momentum-dependent metric is simply of Riemannian type. We can further define the Ricci scalar as usual
\begin{equation}
    S\equiv g_{ij}S_k^{~ikj}.\label{MomRic}
\end{equation}

Unfortunately, the metric which will be treated below does not generally yield a totally symmetric Cartan tensor \eqref{cartan}. Thus, we are dealing with a generalized Hamilton space. In this case, the nonlinear connection must be provided beforehand. By analogy with the simpler case, we choose the nonlinear connection to vanish because the metric harbours no position dependence. Then, the same reasoning follows. 

A note of caution might be in order, though. Have in mind, that the metric still constitutes a tensor and thus transforms as such. It can only be independent of the position if the system is described in Cartesian coordinates. Otherwise, several issues arise which complicate the process of quantization enormously. Fortunately, this set of coordinates suffices for the purpose of the present paper. 

\subsection{Quantum mechanics}\label{subsec:curvqm}

Given a metric \eqref{mommet} and a vanishing nonlinear connection, it is possible to construct the line element in momentum space
\begin{equation}
    \D \sigma^2=g^{ij}(p)\D p_i\D p_j.
\end{equation}
First and foremost, this implies that the dynamics of a single particle derive from a Hamiltonian operator
\begin{equation}
    \hat{H}=\frac{1}{2m}\hp_i\hp_jg^{ij}\left(\hp\right)+V\left(\hx^i\right).
\end{equation}
Furthermore, the position and momentum operators obey the Heisenberg algebra
\begin{align}
    [\hat{x}^i,\hat{x}^j]=0\hspace{.5cm}[\hat{p}_i,\hat{p}_j]=0\hspace{.5cm}[\hat{x}^i,\hat{p}_j]=i\hbar\delta^i_j,\label{HeisAlg}
\end{align}
as in textbook quantum mechanics. 

A convenient representation of this algebra yielding a Hermitian Hamiltonian is based on the integral measure
\begin{equation}
    \D\mu(p)=\D^dp\sqrt{g(p)}\label{curvmeasure}
\end{equation}
with the determinant of the metric $g=\det g^{ij}.$ Then, the Hilbert space scalar product, transforming as a scalar if the momentum space wave functions $\psi$ and $\phi$ transform as scalars, becomes
\begin{equation}
    \Braket*{\psi}{\phi}=\int\D^dp\sqrt{g(p)}\psi^*(p)\phi(p).
\end{equation}
The position operator, being an observable, is required to be symmetric with respect to the measure \eqref{curvmeasure} which is why it is turned into a vertical covariant derivative denoted by the symbol $\dot{\nabla}$
\begin{equation}
    \hat{x}^i\psi=i\hbar\left(\dot{\partial}^i+\frac{1}{2}C^{ij}_j\right)\psi=i\hbar \dot{\nabla}^i\psi.\label{curvedmomspaceposop}
\end{equation}
Correspondingly, the Hamiltonian describing a single particle in curved momentum space acts on wave functions as
\begin{equation}
    \hat{H}\psi=\left[\frac{1}{2m}g^{ij}(p)p_ip_j+V\left(i\hbar\dot{\nabla}^i\right)\right]\psi.\label{curvham}
\end{equation}
Furthermore, the geodesic distance $\sigma$, the only possible position-dependent scalar appearing in the Hamiltonian, can be computed solving the differential equation 
\begin{equation}
    g^{ij}\partial_i\sigma^2\partial_j\sigma^2=4\sigma^2.
\end{equation}
In the given case, this procedure results in the expression
\begin{equation}
    \sigma^2=g_{ij}(p)\left(x-x_0\right)^i\left(x-x_0\right)^j,
\end{equation}
where $x_0^i$ denote the coordinates of the point with respect to which the distance is calculated, for reasons of simplicity chosen to coincide with the origin $x_0^i=0.$ Considering the nonvanishing commutator of positions and momenta, this clearly leads to operator ordering ambiguities analogous to the ones appearing in the kinetic energy of a particle on a curved background. Similarly, they can be resolved representing the squared geodesic distance as the Laplace-Beltrami operator in momentum space
\begin{equation}
    \hat{\sigma}^2\psi=-\hbar^2\frac{1}{\sqrt{g}}\dot{\partial}^i\left(\sqrt{g}g_{ij}\dot{\partial}^j\psi\right),\label{actgeoddist}
\end{equation}
which is clearly Hermitian with respect to the measure \eqref{curvmeasure}. 

Evidently, this description bears much resemblance to quantum mechanics on a spatially curved manifold. Keep in mind, though, that this picture does not hold under general coordinate transformations.

\section{GUP-deformed quantum mechanics}\label{sec:GUPQM}

In contrast to the theory described in the previous section, quantum mechanics with a minimum length is derived from a deformed algebra of observables
\begin{subequations}
\label{GUPalg}
\begin{align}
    [\hat{x}^a,\hat{x}^b]=&i\hbar\tilde{f}^{ab}(\hat{x},\hat{p})\\
    [\hat{p}_a,\hat{p}_b]=&0\\
    [\hat{x}^a,\hat{p}_b]=&i\hbar f^a_b(\hat{p}),\label{posmomgupalg}
\end{align}
\end{subequations}
where we introduced the tensor-valued functions $\tilde{f}^{ab}(\hat{x},\hat{p})$ and $f^a_b(\hp)$ which are not independent. Instead, they are constrained by the Jacobi identity
\begin{equation}
    \left[\tilde{f}^{ab},\hat{p}_c\right]=2\left[f^{[a}_c,\hx^{b]}\right]\label{JacobiId}
\end{equation}
where the square brackets denote antisymmetrization.

The usual way to go at this point consists in finding a representation in momentum space for this algebra. For example, the position operator may read \cite{Kempf94}
\begin{equation}
    \hat{x}^a\psi=i\hbar f^a_b(p)\dot{\partial}^b\psi\label{GUPposop}.
\end{equation}
Within this representation, the Jacobi identity \eqref{JacobiId} can be solved yielding
\begin{equation}
    \tilde{f}^{ab}=2f^{[a}_c\dot{\partial}^{|c|}f^{b]}_d\left(f^{-1}\right)^d_ex^e\propto \hat{J}^{ba}\label{noncom}
\end{equation}
where we introduced the angular momentum operator $\hat{J}^{ab}=2\hx^{[a}\hp^{b]}.$ 

At first glance, the theory of generalized uncertainty principles and the theory of curved momentum space differ substantially. How, then, can they be reconciled with each other?

\section{Equivalence of the modifications}\label{sec:equivalence}

The algebra \eqref{GUPalg} indicates that the kinematical description in the GUP approach is based on unusual coordinates in phase space. In particular, they are not of Darboux-form which would imply the canonical commutation relations \eqref{HeisAlg} to be satisfied. The Darboux theorem \cite{Darboux82}, however, states that symplectic manifolds, like phase space, have vanishing curvature. Thus, provided the necessary transformation is found, every system can be expressed in terms of Darboux coordinates. The task of this section entails finding new operators 
\begin{equation}
    \hat{x}^a\rightarrow \hat{X}^i\left(\hat{x},\hat{p}\right)\hspace{1cm} \hat{p}_a\rightarrow\hat{P}_i\left(\hat{p}\right),\label{pstrans}
\end{equation}
such that $\hat{X}^i$ and $\hat{P}_i$ satisfy the Heisenberg algebra \eqref{HeisAlg}. A similar approach albeit with different realization and goals was followed in Ref. \cite{Galan07} in the context of doubly special relativity.

\subsection{Transformation}

Let us, in particular, assume that the transformation take the shape
\begin{subequations}\label{trans}
\begin{align}
    \hx^a=&\left(e^{-1}\right)^a_i(\hP)\hX^i\label{xtrans}\\
    \hp_a=&e_a^i(\hP)\hP_i,\label{ptrans}
\end{align}
\end{subequations}
where the coordinates transform according to the operator ordering imposed by geometric calculus \cite{Pavsic01} applied to momentum space. Note that other operator orderings would yield equivalent theories \cite{Bosso20,Bosso21} which, however, would not manifestly unveil the nontriviality of momentum space. 

This transformation immediately implies that the Hamiltonian describing the dynamics of a nonrelativistic particle may be reexpressed as
\begin{equation}
    \hat{H}=\frac{1}{2m}\hP_i\hP_je^i_ae^j_b\delta^{ab}+V\left[\left(e^{-1}\right)^a_i\hX^i\right].\label{hamiltaftertrafo}
\end{equation}
Moreover, the geodesic distance in the original flat background transforms in a similar way to the kinetic energy
\begin{equation}
\hat{\sigma}^2=\delta_{ab}\hx^a\hx^b=\delta_{ab}\left(e^{-1}\right)^a_i\hX^i\left(e^{-1}\right)^b_j\hX^j.
\end{equation} 
Thus, the matrix $e^i_a$ may be understood as \emph{vielbein}. Then, we may construct the metric and its inverse as
\begin{align}
    g^{ij}=&\delta^{ab}e^i_ae^j_b\label{metfromviel}\\
    g_{ij}=&\delta_{ab}\left(e^{-1}\right)_i^a\left(e^{-1}\right)_j^b.
\end{align}
Correspondingly, the Hamiltonian acts in momentum space as
\begin{align}
    \hat{H}\psi(P)=& \frac{P_i P_jg^{ij}}{2m}\psi(P)
    + V\left[i\hbar\left(e^{-1}\right)^a_i\dot{\partial}^i\right]\psi(P),\label{HamAfterTrafo}
\end{align}
while the geodesic distance exactly follows Eq. \eqref{actgeoddist}. For this structure to be consistent, the measure has to read
\begin{equation}
\D\mu=\det \left(e^i_a\right)\D^dp,
\end{equation} 
\ie represent the volume form derived from the metric.

Under the assumption, that the transformed phase space coordinates obey the Heisenberg algebra, the commutator of positions and momenta \eqref{GUPalg} implies the Jacobian
\begin{equation}
    \dot{\partial}^aP_j=\left(f^{-1}\right)_b^a\left(e^{-1}\right)_j^b,\label{Jacobian}
\end{equation}
which may be rewritten as a condition on the \emph{vielbein}
\begin{equation}
    f^{[a}_d\left[\dot{\partial}^{|d|}\left(e^{-1}\right)^{b]}_je^j_c-\dot{\partial}^{|d|}f^{b]}_d\left(f^{-1}\right)^d_c\right]=0.\label{conscondgen1}
\end{equation}
Then, after some algebra the tensor measuring the spatial noncommutativity reads
\begin{equation}
	\tilde{f}^{ab}=2f_c^{[a}\dot{\partial}^{|c|}f^{b]}_d\left(f^{-1}\right)^d_ex^e.\label{conscondgen2}
\end{equation}
Fortunately, this relation, derived from the assumptions that the new phase space coordinates obey the Heisenberg algebra and that the original variables satisfy the commutation relations \eqref{GUPalg} and the Jacobi identity \eqref{JacobiId}, reproduces the condition on the noncommutativity of space in the original representation \eqref{noncom}. Thus, the transformation introduced in this paper can always be performed.

To put it in a nutshell, it is possible to describe the dynamics implied by any set of deformed commutators of the form \eqref{GUPalg} by Darboux coordinates defined in Eqs. \eqref{xtrans} and \eqref{ptrans} if the matrix characterizing the transition satisfies the consistency condition \eqref{conscondgen1} and the noncommutativity of the spatial coordinates is of the form \eqref{conscondgen2}. The background, which the system is moving on, will then necessarily be nontrivial.

Note, though, that this is how the metric can be determined in terms of the original momenta $\hp_a.$ In principle, as can be seen from the equation
\begin{equation}
    e^i_a\left(\hp_b\right)=e^i_a\left[e^j_b\left(\hp_c\right)\hP_j\right]=\dots,\label{infreg}
\end{equation}
trying to express the result in terms of the transformed momenta $\hP_i,$ leads to an infinite regress. Yet, this problem can be circumvented by solving it iteratively as in perturbation theory. Before  we get to this point, though, it is instructive to show how the consistency conditions turn out when $\tilde{f}^{ab},$ $f^a_b$ and $e^i_a$ are expressed in terms of scalar functions.

\subsection{Conditions on scalars}

As may be deduced from the Jacobi identity \eqref{JacobiId}, the spatial noncommutativity depends on the original phase space variables as
\begin{equation}
	 \tilde{f}^{ab}=\tilde{f}\left(\hp^2\right)\hat{J}^{ba},
\end{equation}
where the newly introduced dimensionful scalar $\tilde{f}$ measures the noncommutativity of space. Furthermore, expressed in a way similar to Refs. \cite{Chang10,Chang11,Chang16}, the quantity $f^a_b,$ being a tensor, assumes the form
\begin{equation}
    f^a_b=A\left(\hp^2\right)\delta^a_b+B\left(\hp^2\right)\frac{\hp^a\hp_b}{\hat{p}^2},\label{introAB}
\end{equation}
where we introduced the dimensionless scalars $A$ and $B.$ Note that they have to satisfy the conditions $A(0)=1$ and $B(0)=0$ for the given phase space variables to reduce to ordinary canonical conjugates in the low-energy limit. Both scalars are related to the function $\tilde{f}$ according to Eq. \eqref{conscondgen2}
\begin{equation}
    \tilde{f}=2\left(\log A\right)'(A+B)-\frac{B}{\hp^2},\label{noncomcondsca}
\end{equation}
where the prime denotes derivation with respect to $\hp^2.$ 

Furthermore, providing the \emph{vielbein} in the most general form compatible with the generalized uncertainty principle
\begin{equation}
    e^i_a=C\left(\hp^2\right)\delta^i_a+D\left(\hp^2\right)\frac{\hp^i\hp_a}{\hp^2},\label{introCD}
\end{equation}
Eq. \eqref{conscondgen1} suffices to determine the newly introduced dimensionless scalar functions $C$ and $D$ implying the relation
\begin{equation}
    \frac{D}{C}=\left[\tilde{f}+2\left(\log C\right)'(A+B)\right]\hp^2=A-1,\label{conscondgen2res}
\end{equation}
which, assuming that the background reduces to flat space in the low-energy limit, \ie $C(0)=1$ and $D(0)=0,$ can be solved to yield
\begin{align}
    C=&\exp\left(\frac{1}{2}\int_0^{\hp^2}\frac{A-1-\tilde{f}}{A+B}(q)\D q\right)\label{Cres}\\
    D=&(A-1)C\label{Dres}.
\end{align}

Have in mind, though, that the expression for the \emph{vielbein} \eqref{introCD} needs to be translated to a description in terms of the canonical momenta in accordance with Eq. \eqref{infreg}. The metric can then be obtained from Eq. \eqref{metfromviel} as
\begin{equation}
    g^{ij}=C^2\delta^{ij}+\left(2CD+D^2\right)\hP^i\hP^j.
\end{equation}
In short, we can understand the generalized uncertainty principle as dual description to a quantum theory on nontrivial momentum space. Additionally, the newly found set of phase space variables allows for applications in its own right.

\subsection{Note on canonical variables}

Classically, the dynamics of any system are governed by the action describing it. Alternatively, in quantum theory it suffices to provide a Hamiltonian and an algebra relating the dynamical variables. In the Heisenberg picture, the evolution of the system may then be obtained according to the Heisenberg equations. To provide the corresponding Schrödinger equation and the action of a system, however, it is compulsory to find canonically conjugate variables, \ie a set obeying the Heisenberg algebra \eqref{HeisAlg}. By construction, this is the case considering the phase space coordinates introduced in the preceding section  \eqref{trans}. Furthermore, it is evident that the Heisenberg equations of motion in terms of both sets provided in this paper are equivalent. Thus, the action of the system, subject to a generalized uncertainty principle including spatial noncommutativity, reads
\begin{align}
    S=&\int\D t\left[\dot{X}^iP_i-H(X,P)\right].
\end{align}
Up until now, this kind of result had only been obtained in the case of a commutative space \cite{Brau06,Bosso18b} which is related to the one provided in the present paper by a canonical transformation.

\subsection{Iterative approach}

For all intents and purposes, it suffices to solve Eqs. \eqref{noncomcondsca} and \eqref{conscondgen2res} iteratively. Assume as given the coefficients of a power series expansion of $A$ and $B$
\begin{equation}
    A=\sum_nA_n\left(\frac{l\hp}{\hbar}\right)^{2n} \hspace{.5cm} B=\sum_nB_n\left(\frac{l\hp}{\hbar}\right)^{2n}
\end{equation}
with some length scale $l$ and where $B_0=0$ to avoid divergences. Similarly, describe the scalars $\tilde{f},$ $C$ and $D$ using power series
\begin{align}
    \tilde{f}=&\frac{1}{\hp^2}\sum_n\tilde{f}_{n-1}\left(\frac{l\hp}{\hbar}\right)^{2n}\\ 
    C=&\sum_nC_n\left(\frac{l\hp}{\hbar}\right)^{2n}\\ 
    D=&\sum_nD_n\left(\frac{l\hp}{\hbar}\right)^{2n},
\end{align}
where now $D_0=\tilde{f}_{-1}=0.$ Then, Eq. \eqref{noncomcondsca} becomes at $N$th order
\begin{equation}
    \sum_{n=0}^NA_{N-n}\left[2(N-n)\left(A_n+B_n\right)-B_n-\tilde{f}_{n}\right]=0
\end{equation}
determining the coefficients $f_n$ order by order. Moreover, the Eqs. \eqref{conscondgen2res} uniquely specify the dependence of the coefficients $C_n$ and $D_n$ on $A_n$ and $B_n$ in an analogous fashion
\begin{align}
    D_N=&\sum_{n=0}^NC_{N-n}\left[\tilde{f}_{n}+2(N-n)\left(A_{n}+B_{n}\right)\right]\\
    =&\sum_{n=0}^NC_{N-n}A_n-C_N.
\end{align}

In short, the coefficients of the power series expansions describing the functions $C$ and $D$ are related to the ones representing the given scalars $A$ and $B$ such that there is no ambiguity. This opens up the possibility for a perturbative treatment.

\subsection{Application to the quadratic generalized uncertainty principle}\label{sec:applquadGUP}

As mentioned above, under the assumption that the generalized uncertainty principle recover Heisenberg's relation in the low-energy limit, the unperturbed scalars have to satisfy $A_0=1$ and $B_0=\tilde{f}_{-1}=0.$ Furthermore, denote $A_1=\beta,$ $B_1=\beta'$ and choose the Planck length to describe the scale to compare to ($l=l_p$) in accordance with the literature \cite{Das09,Tawfik14,Hossenfelder12}. Accordingly, we find
\begin{align}
    \tilde{f}_0=&0&\tilde{f}_1=&2\beta-\beta'\\
    C_0=&1&C_1=&\frac{\beta'-\beta}{2}\\
    D_0=&0&D_1=&\beta.
\end{align}
At second order, the contribution stemming from the iterative apperance of the \emph{vielbein} \eqref{infreg} is trivial. Thus, the metric reads
\begin{equation}
    g^{ij}=\delta^{ij}+h^{ij},
\end{equation}
where the correction to the Euclidean part results as
\begin{equation}
    h^{ij}=\left(\beta'-\beta\right)\left(\frac{l_p\hP}{\hbar}\right)^2\delta^{ij}+2\beta\left(\frac{l_p}{\hbar}\right)^2\hP^i\hP^j.\label{metpert}
\end{equation}
Hence, we can derive the Cartan tensor from it yielding
\begin{equation}
    C^{ijk}=2\left(\frac{l_P}{\hbar}\right)^2\left[\left(\beta'-\beta\right)\hP^i\delta^{jk}+2\beta \hP^{(j}\delta^{k)i}\right]\label{GUPcartan}.
\end{equation}
The Cartan tensor is totally symmetric if and only if $\beta'=2\beta,$ \ie $\tilde{f}\simeq 0,$ implying a commutative background. Then, the scalar \eqref{cartsca} derived from it reads in the low-energy limit
\begin{equation}
    \left.C\right|_{\hP=0}=2d(d+2)\beta\left(\frac{l_p}{\hbar}\right)^2.\label{Cquad}
\end{equation}
Otherwise, this metric does not belong to the class of Hamilton spaces as claimed in section \ref{subsec:curvqm}. Nevertheless assuming a vanishing nonlinear connection as was argued in the same section, the curvature tensor in momentum space \eqref{MomRiem} can be determined. In the low-energy limit it reads
\begin{equation}
    \left.S^{ikjl}\right|_{\hP=0}=2\tilde{f}_1\left(\frac{l_p}{\hbar}\right)^2\left(\delta^{ij}\delta^{kl}-\delta^{il}\delta^{kj}\right).\label{GUPRiem}
\end{equation}
Given this result, it is possible to compute the Ricci scalar in accordance with Eq. \eqref{MomRic}
\begin{equation}
    \left.S\right|_{\hP=0}=2d(d-1)\tilde{f_1}\left(\frac{l_p}{\hbar}\right)^2.\label{Squad}
\end{equation}
Thus, at first order the curvature of momentum space, provided the system is represented canonically, measures the noncommutativity of space described in terms of the original coordinates. This is why the Cartan tensor is totally symmetric in the case of a GUP with a commutative background. Note, though, that, despite the background being flat, the momentum basis in terms of which the system is hence described is not the usual one. As the symplectic structure is not invariant under nonlinear transformations of momenta, the resulting theory is not equivalent to ordinary quantum mechanics notwithstanding the flat background. This effect is measured by the quantity $C$ \eqref{Cquad}.

In short, quadratically deformed Heisenberg algebras may be understood as a normal-frame-description of a momentum space harbouring essentially Planckian curvature if space is noncommutative. Thus, we can import much information from the phenomenology of generalized uncertainty principles to this arena.

\section{Constraints from existing literature}\label{sec:constraints}

In the preceding section, a correspondence between models of the quadratic generalized uncertainty principle and quantum mechanics on a non-Euclidean momentum space was pointed out. This connection implies that bounds on the noncommutativity of space $\tilde{f}_1$ immediately carry over to the curvature tensor in momentum space in accordance with equation \eqref{GUPRiem}. Some of these, mostly extracted from Ref. \cite{Hinchliffe02}, are displayed in table \ref{tab:labgupS}. The dominating constraint on the curvature scalar \eqref{Squad} stems from the dipole moment of the electron yielding
\begin{equation}
    \left.S\right|_{p=0}<10^{27}m_p^{-2}.
\end{equation}
Note that spatially noncommutative geometry may lead to direct violations of Lorentz-invariance \cite{Anisimov01}, which would push this bound into the Planckian regime. However, depending on the relativistic generalization of the model, the symmetry might only be deformed yielding much weaker constraints.

\begin{table}[]
    \centering
\begin{tabular}{ c c || c } 
 Experiment & Ref.  & Upper bound on $Sm_p^2$\\
 \hline
 electron dipole moment & \cite{Hinchliffe01} & $10^{27}$\\
 lamb shift & \cite{Chaichian00a,Chaichian02} & $10^{29}$\\
 $^9$Be decay & \cite{Carroll01} & $10^{29}$\\ 
 composite quarks/ leptons & \cite{Ghoderao18,Tanabashi18} & $10^{29}$\\
 M\o ller scattering & \cite{Hewett00} & $10^{31}$\\
 muon $g-2$ & \cite{Kersting01} & $10^{31}$\\
 hydrogen spectrum & \cite{Gnatenko14,Akhoury03} & $10^{33}$\\
 $^{133}$Cs decay & \cite{Carroll01} & $10^{35}$\\
 star energy loss & \cite{Schupp02} & $10^{35}$\\
 Pauli oscillator & \cite{Heddar21} & $10^{41}$\\ 
 Aharonov-Bohm & \cite{Chaichian00b} & $10^{43}$
\end{tabular}
\caption{Upper bounds on the low-energy-limit of the scalar curvature in momentum space as in Eq. \eqref{Squad} given in units of $l_p^2/\hbar^2= m_p^{-2}.$ \label{tab:labgupS}}
\end{table}

Furthermore, in the case of a commutative background space ($\tilde{f}_1=0$) bounds on the parameter $\beta$ can be translated as limits to the deviation from the usual momentum basis embodied by the scalar $C$ \eqref{Cquad}. A selection of bounds obtained this way is on display in table \ref{tab:labgupC}. Note here, that experiments involving pendula \cite{Bawaj14}, harmonic oscillators \cite{Bushev19} and optomechanical setups \cite{Khodadi17} deal with macroscopic quantum objects. As there are reservations towards the direct adoption of results from multiparticle states to the mechanics of single particles (see \eg Ref. \cite{Amelino-Camelia13}), those should be taken with a grain of salt. The strongest constraint excluding macroscopic experiments is derived from the anomalous magnetic moment of the muon \cite{Das11} implying that
\begin{equation}
    \left.C\right|_{p=0}<10^{17}m_P^{-2}.\label{CConst}
\end{equation}

Summarizing, both the curvature of momentum space as well as the deviation from the canonical momentum basis in the flat case are constrained experimentally from bounds on the noncommutativity of space and on the $\beta$-parameter of the commutative quadratic generalized uncertainty principle, respectively.

\begin{table}[]
    \centering
\begin{tabular}{ c c || c } 
 Experiment & Ref.  & Upper bound on $Cm_p^2$\\
 \hline
 pendula & \cite{Bawaj14} & $10^5$\\
 harmonic oscillators & \cite{Bushev19,Bawaj14} & $10^8$\\ 
 muon $g-2$  & \cite{Das11} & $10^{17}$\\
 equivalence principle & \cite{Ghosh13} & $10^{20}$\\ 
 quantum noise & \cite{Girdhar20} & $10^{22}$\\
 tunnelling microscope & \cite{Das08} & $10^{22}$\\
 hydrogen spectrum & \cite{Bouaziz10,AntonacciOakes13} & $10^{23}$\\
 gravitational bar detectors & \cite{Marin13,Marin14} & $10^{33}$\\ 
 lamb shift & \cite{Das08,Ali11b} & $10^{37}$\\ 
 $^{87}$Rb interferometry & \cite{Gao16,Khodadi18a} & $10^{40}$
\end{tabular}
\caption{Upper bounds on the deviation from the canonical basis in momentum space $C$ as in Eq. \eqref{Cquad} given in units of $l_p^2/\hbar^2= m_p^{-2}.$ \label{tab:labgupC}}
\end{table}

\section{Conclusion}\label{sec:Conclusion}

Modifications to the Heisenberg algebra yield a convenient way to incorporate minimum length effects, a generic prediction of quantum gravity, into nonrelativistic quantum mechanics. Recent results \cite{Schuermann18,Dabrowski19,Dabrowski20,Petruzziello21} suggest a deep connection between such generalized and extended uncertainty principles and non-Euclidean momentum and position spaces, respectively. In this paper we further strengthened this connection presenting a noncanonical transformation which provides a direct map from theories involving generalized uncertainty principles to quantum mechanics on curved momentum space.

In that vein, we first introduced quantum mechanics on a background described by a purely momentum-dependent metric. We further gave an account of the kind of general changes to the canonical commutation relations which are usually associated to generalized uncertainty relations including noncommutativity of the position coordinates. Bringing those two lines of thought together, we found an explicit dual description of this type of deformation in terms of a nontrivial momentum space. In other words, every generalized uncertainty principle entailing a certain set of non-Darboux coordinates yields its counterpart in a specific set of canonically conjugated phase space variables. The resulting dynamics strongly indicate the presence of a nontrivial momentum space.

In particular, in the case of the quadratic generalized uncertainty principle the curvature tensor in momentum space is proportional to the spatial noncommutativity. However, the dual description of a commutative space does not imply a trivial background because the corresponding basis in momentum space is curvilinear. As nonlinear basis transformations in momentum space are not canonical, the resulting theory is inequivalent to ordinary quantum mechanics. The deviation from Riemannian geometry induced by this unusual basis can then be measured by a scalar derived from the Cartan tensor.

This allows us to import constraints on the curvature of momentum space from bounds on the noncommutativity of space yielding for the Ricci scalar in momentum space $S|_{p=0}<10^{13}m_p^{-2}.$ Moreover, the literature on generalized uncertainty principles with commutative space is helpful in constraining the deviation from Riemannian geometry when the curvature is vanishing yielding $C_{p=0}<10^{17}m_p^2.$

Evidently, the reasoning applied in the present paper is general enough to be applied to extended uncertainty principles in an analogous fashion. Correspondingly, those can be mapped to theories of quantum mechanics on curved position space thus establishing the connection hinted at in Refs. \cite{Schuermann18,Dabrowski19,Dabrowski20,Petruzziello21}. 

To make a long story short, the interplay of generalized uncertainty principles and non-Euclidean momentum space as well as extended uncertainty principles and curved position space yields a rich phenomenology that justifies further investigation. In particular, a formulation of quantum mechanics on generalized Hamilton spaces away from Cartesian coordinates, such that the metric may depend on positions and momenta, may be seen as a goal to achieve in future work. 
\section*{Acknowledgments}

The author thanks Christian Pfeifer and Samuel Barroso-Bellido for useful discussions and is further indebted to Pasquale Bosso for helpful comments. His work was supported by the Polish National Research and Development Center (NCBR) project ''UNIWERSYTET 2.0. --  STREFA KARIERY'', POWR.03.05.00-00-Z064/17-00 (2018-2022). Moreover, the author would like to acknowledge the contribution of the COST Action CA18108.

\bibliographystyle{unsrt}
\bibliography{bib}

\end{document}